\documentstyle[12pt]{article}
\catcode`\@=11
\long\def\@makefntext#1{
\protect\noindent \hbox to 3.2pt {\hskip-.9pt
$^{{\ninerm\@thefnmark}}$\hfil}#1\hfill}
\def\@makefnmark{\hbox to 0pt{$^{\@thefnmark}$\hss}}
\def\ps@myheadings{\let\@mkboth\@gobbletwo
\def\@oddhead{\hbox{}\rightmark\hfil\ninerm\thepage}
\def\@oddfoot{}\def\@evenhead{\ninerm\thepage\hfil
\leftmark\hbox{}}\def\@evenfoot{}
\def\sectionmark##1{}\def\subsectionmark##1{}}
\newcounter{sectionc}\newcounter{subsectionc}\newcounter{subsubsectionc}
\renewcommand{\section}[1] {\vspace{0.6cm}\addtocounter{sectionc}{1}
\setcounter{subsectionc}{0}\setcounter{subsubsectionc}{0}\noindent
{\bf\thesectionc. #1}\par\vspace{0.4cm}}
\renewcommand{\subsection}[1] {\vspace{0.6cm}\addtocounter{subsectionc}{1}
\setcounter{subsubsectionc}{0}\noindent
{\it\thesectionc.\thesubsectionc. #1}\par\vspace{0.4cm}}
\def\abstracts#1{{
\centering{\begin{minipage}{30pc}\tenrm\baselineskip=12pt\noindent
\centerline{\tenrm ABSTRACT}\vspace{0.3cm}
\parindent=0pt #1\end{minipage}}\par}}
\renewenvironment{thebibliography}[1]{\begin{list}{\arabic{enumi}.}
{\usecounter{enumi}\setlength{\parsep}{0pt}
\setlength{\leftmargin 1.25cm}{\rightmargin 0pt}
\setlength{\itemsep}{0pt} \settowidth {\labelwidth}{#1.}\sloppy}}{\end{list}}
\def\@citex[#1]#2{\if@filesw\immediate\write\@auxout {\string\citation{#2}}\fi
\def\@citea{}\@cite{\@for\@citeb:=#2\do {\@citea\def\@citea{,}\@ifundefined
{b@\@citeb}{{\bf ?}\@warning
{Citation `\@citeb' on page \thepage \space undefined}}
{\csname b@\@citeb\endcsname}}}{#1}}
\newif\if@cghi
\def\cite{\@cghitrue\@ifnextchar [{\@tempswatrue
\@citex}{\@tempswafalse\@citex[]}}
\def\@cite#1#2{{$\null^{#1}$\if@tempswa \fi}}
\begin{document}
\begin{titlepage}
\large
\hfill\begin{tabular}{l}HEPHY-PUB 610/94\\ UWThPh-1994-43\\ September 1994
\end{tabular}\\[2cm]
\begin{center}
{\Large\bf ALL AROUND}\\[.5ex]
{\Large\bf THE SPINLESS SALPETER EQUATION}\\
\vspace{1.5cm}
{\Large\bf Wolfgang LUCHA}\\[.5cm]
Institut f\"ur Hochenergiephysik,\\
\"Osterreichische Akademie der Wissenschaften,\\
Nikolsdorfergasse 18, A-1050 Wien, Austria\\[1cm]
{\Large\bf Franz F.~SCH\"OBERL}\\[.5cm]
Institut f\"ur Theoretische Physik,\\
Universit\"at Wien,\\
Boltzmanngasse 5, A-1090 Wien, Austria\\[1.5cm]
\normalsize\it
Invited talk presented by F.~Sch\"oberl at the workshop\\
``QUARK CONFINEMENT AND THE HADRON SPECTRUM''\\
Como, Italy, June 20--24, 1994\\
To appear in the Proceedings
\end{center}
\vspace{1.5cm}
\abstracts{We review some important topics related to the semirelativistic
description of bound states by the spinless Salpeter equation: the special
case of the Coulomb interaction, numerical approximation methods, and a way
to avoid the problematic square-root operator of the relativistic kinetic
energy.}
\end{titlepage}

\rm\baselineskip=14pt

\section{Introduction}

The ``spinless Salpeter equation'' represents a well-defined standard
approximation to the Bethe--Salpeter formalism for the description of bound
states within relativistic quantum field theories. It may be derived from the
Bethe--Salpeter equation\cite{salpeter51}
\begin{enumerate}
\item by eliminating, in full accordance with the spirit of an instantaneous
interaction, any dependence on timelike variables, which leads to the
Salpeter equation,\cite{salpeter52} and
\item by neglecting any reference to the spin degrees of freedom of the
involved bound-state constituents and restricting exclusively to
positive-energy solutions.
\end{enumerate}
The Hamiltonian governing the dynamics of the quantum system under
consideration
\begin{itemize}
\item incorporates relativistic kinematics by involving the square-root
operator of the relativistic kinetic energy, $\sqrt{{\bf p}^2 + m^2}$ for
particles of mass $m$ and momentum ${\bf p}$, but
\item describes the forces acting between the bound-state constituents by an
arbitrary coordinate-dependent static interaction potential $V({\bf x})$.
\end{itemize}
For the case of bound states consisting of two particles of equal mass $m$,
the generic Hamiltonian $H$ in the center-of-momentum frame of these
constituents, expressed in terms of the relative momentum ${\bf p}$ and the
relative coordinate ${\bf x}$, reads
\begin{equation}
H = 2\sqrt{{\bf p}^2 + m^2} + V({\bf x}) \ .
\label{eq:semrelham}
\end{equation}
The equation of motion involving this type of Hamiltonian, $$H|\psi\rangle =
E|\psi\rangle\ ,$$ with energy eigenvalues $E$ and corresponding
Hilbert-space eigenstates $|\psi\rangle$, has been widely used, e.~g., for
the (semi-)relativistic description of hadrons as bound states of quarks
within the framework of potential models.\cite{lucha91,lucha92,lucha92sign}

\section{The Spinless Relativistic Coulomb Problem}

Without doubt, a central r\^{o}le in physics is played by the Coulomb
potential $V_{\rm C}(r)$. This Coulomb potential is a spherically symmetric
potential, i.~e., one which depends only on the radial coordinate $r \equiv
|{\bf x}|$; its interaction strength is parametrized by some coupling
constant $\kappa$:
\begin{equation}
V({\bf x}) = V_{\rm C}(r) = - \frac{\kappa}{r} \ , \quad \kappa > 0 \ .
\label{eq:coulpot}
\end{equation}
The bound-state problem defined by the semirelativistic Hamiltonian of
Eq.~(\ref{eq:semrelham}) with the Coulomb potential (\ref{eq:coulpot}) is
what we call the ``spinless relativistic Coulomb problem.''

\subsection{(Mostly) Analytic Results}
\vspace{-0.35cm}

Over the past years, the spinless relativistic Coulomb problem has been the
subject of intense study. We summarize, in chronological order, the knowledge
gained so far:
\begin{itemize}
\item Herbst,\cite{herbst77} in a rigorous mathematical discussion, developed
the complete spectral theory of the one-particle counterpart of the operator
(\ref{eq:semrelham}), (\ref{eq:coulpot})---from which one may directly deduce
for the two-particle semirelativistic Coulombic Hamiltonian under
consideration its essential self-adjointness for $\kappa \le 1$ and the
existence of its Friedrichs extension up to some critical value, viz.,
$$\kappa_{\rm cr} = \frac{4}{\pi}\ ,$$ of the coupling constant---and derived
some strict lower bound on the energy $E_0$ of the ground state which
translates in the two-particle case to
$$
E_0 \ge 2\,m\sqrt{1 - \left(\frac{\pi\,\kappa}{4}\right)^2}
\quad \mbox{for} \ \kappa < \frac{4}{\pi} \ .
$$
\item Durand and Durand\cite{durand83} presented in an involved analysis of
the spinless relativistic Coulomb problem for the particular case of
vanishing orbital angular momentum of the bound-state constituents the
explicit construction of the analytic solution for the corresponding wave
function.
\item Castorina {\it et~al.}\cite{castorina84} generalized the behaviour of
this wave function near the origin, that is, for small relative distances of
the bound-state constituents, to arbitrary values of the orbital angular
momentum.
\item Hardekopf and Sucher\cite{hardekopf85} carried out a comprehensive
numerical investigation of one- and two-particle relativistic wave equations
for both spin-$0$ as well as spin-$\frac{1}{2}$ particles. Their numerical
results for the ground-state energy $E_0$ of the spinless relativistic
Coulomb problem are consistent with the lower bound of Herbst and provide no
evidence that this ground-state energy indeed approaches zero as the coupling
constant $\kappa$ rises to its critical value $\kappa_{\rm cr} = 4/\pi$, that
is, $$E_0(\kappa = \kappa_{\rm cr}) \ne 0\ .$$
\item Martin and Roy\cite{martin89} improved the lower bound of Herbst
somewhat to
$$
E_0 \ge 2\,m\sqrt{\frac{1 + \sqrt{1 - \kappa^2}}{2}}
\quad \mbox{for} \ \kappa < 1 \ .
$$
\item Raynal {\it et~al.}\cite{raynal94} made use of a generalized version of
the ``local-energy'' theorem:\cite{barnsley78} Assume
\begin{enumerate}
\item that the Fourier transform $\tilde V({\bf p})$ of the interaction
potential $V({\bf x})$ is strictly negative, except at infinity, as is the
case for the Coulomb potential (\ref{eq:coulpot}),
\item that the spectrum of the Hamiltonian under consideration, $H$, is
discrete, and
\item that the ground state of the Hamiltonian exists.
\end{enumerate}
Define the local energy $${\cal E}({\bf p}) \equiv T({\bf p}) + \frac{\int
d^3q\,\tilde V({\bf p} - {\bf q})\,\phi({\bf q})}{\phi({\bf p})}\ ;$$ here
$T({\bf p})$ represents the kinetic energy, which in our case is, of course,
given by $$T({\bf p}) = 2\sqrt{{\bf p}^2 + m^2}\ ,$$ and $\phi({\bf p})$
denotes some suitably chosen, positive trial function, $$\phi({\bf p}) > 0\
.$$ Then the lowest-lying eigenvalue of the Hamiltonian $H$, $E_0$, is
bounded by
$$
\inf_{{\bf p}} {\cal E}({\bf p}) \le E_0 < \sup_{{\bf p}} {\cal E}({\bf p})
\ .
$$
With the help of this theorem, Raynal {\it et~al.}\ succeeded in restricting
numerically the ground-state energy eigenvalue of the semirelativistic
Hamiltonian (\ref{eq:semrelham}), (\ref{eq:coulpot}), considered as a
function of the coupling strength $\kappa$, to some remarkably narrow band.
In particular, at the critical value $\kappa_{\rm cr} = 4/\pi$ of this
coupling constant $\kappa$ they found $$0.4825 < \frac{E_0(\kappa =
\kappa_{\rm cr})}{2\,m} < 0.4843\ ,$$ which proved the nonvanishing of the
ground-state energy eigenvalue $E_0$ at the critical coupling constant.
\item Le Yaouanc {\it et~al.}\cite{leyaouanc94} gave the systematic series
expansion of the eigenvalues of the Hamiltonian (\ref{eq:semrelham}),
(\ref{eq:coulpot}) in powers of the coupling constant $\kappa$ up to and
including the order $O(\kappa^7)$ for arbitrary states with vanishing orbital
angular momentum. For the ground-state energy $E_0$ this expansion reads
$$
\frac{E_0}{2\,m} = 1 - \frac{\kappa^2}{8} - \frac{5\,\kappa^4}{128}
+ \frac{\kappa^5}{12\,\pi} + \frac{\kappa^6}{64}\ln\kappa + O(\kappa^6) \ .
$$
Higher orders of this expansion in the form given by Le Yaouanc {\it et~al.},
however, involve explicitly the corresponding wave functions and their
derivatives, which renders difficult to derive any profit from it. Up to
order $O(\kappa^4)$ the above result coincides with the improved lower bound
of Martin and Roy.
\end{itemize}
The presence of non-analytic $\ln\kappa$ terms in the expansion quoted by Le
Yaouanc {\it et~al.}\ gives a clear hint at the non-analytic nature to be
expected for the energy eigenvalues of the spinless relativistic Coulomb
problem as functions of the coupling strength $\kappa$.

\newpage
\subsection{Analytic Upper Bounds}
\vspace{-0.35cm}

In view of the fact that exact solutions to the spinless relativistic Coulomb
problem are still lacking, we seek for an analytic upper bound on the
ground-state energy level of the semirelativistic Coulombic Hamiltonian
(\ref{eq:semrelham}), (\ref{eq:coulpot}). To this end, we make use of a
rather standard variational technique, which proceeds along the steps of the
following, extremely simple recipe:\cite{lucha92}
\begin{enumerate}
\item Choose a suitable set of trial states $\{|\lambda\rangle\}$. The
different members of this set are distinguished from each other by some sort
of variational parameter $\lambda$.
\item Compute the set of expectation values of the Hamiltonian under
consideration, $H$, with respect to these trial states $|\lambda\rangle$ in
order to obtain $$E(\lambda) \equiv \langle\lambda|H|\lambda\rangle\ .$$
\item Determine, from the first derivative, that value $\lambda_{\rm min}$ of
the variational parameter $\lambda$ which minimizes the resulting,
$\lambda$-dependent expression $E(\lambda)$.
\item Compute $E(\lambda)$ at the point of the minimum $\lambda_{\rm min}$ to
find in this way the minimal expectation value $E(\lambda_{\rm min})$ of the
Hamiltonian $H$ in the Hilbert-space subsector of the chosen trial states
$|\lambda\rangle$.
\end{enumerate}
This minimum $E(\lambda_{\rm min})$ provides, of course, only an upper bound
to the proper energy eigenvalue $E$ of the Hamiltonian $H$: $$E \le
E(\lambda_{\rm min})\ .$$

Let us apply the above simple recipe to the case of the Coulomb potential,
Eq.~(\ref{eq:coulpot}). For the Coulomb potential, the most reasonable choice
of trial states is obviously the one for which the coordinate-space
representation $\psi({\bf x})$ of the states $|\lambda\rangle$ for vanishing
radial and orbital angular momentum quantum numbers is given by the
hydrogen-like trial functions
$$
\psi({\bf x}) = \sqrt{\frac{\lambda^3}{\pi}}\,\exp(- \lambda\,r)
\ , \quad \lambda > 0 \ .
$$
For this particular set of trial functions we obtain for the expectation
values we shall be interested in, namely, the ones of the square of the
momentum ${\bf p}$ and of the inverse of the radial coordinate $r$,
respectively, evaluated with respect to the trial states $|\lambda\rangle$:
$$\langle\lambda|{\bf p}^2|\lambda\rangle = \lambda^2$$ and
$$\left\langle\lambda\left|\frac{1}{r}\right|\lambda\right\rangle = \lambda\
.$$

Now, as an immediate consequence of the fundamental postulates of any quantum
theory, the expectation value of some given Hamiltonian $H$ taken with
respect to any normalized Hilbert-space state and, therefore, in particular,
taken with respect to any of the above trial states must necessarily be
larger than or equal to that eigenvalue $E_0$ of the Hamiltonian $H$ which
corresponds to its ground state: $$E_0 \le E(\lambda) \equiv
\langle\lambda|H|\lambda\rangle\ .$$ The application to the semirelativistic
Hamiltonian of Eq.~(\ref{eq:semrelham}) yields for the right-hand side of
this inequality
\begin{equation}
E(\lambda) =
2\left\langle\lambda\left|\sqrt{{\bf p}^2 + m^2}\right|\lambda\right\rangle
+ \langle\lambda|V({\bf x})|\lambda\rangle \ .
\label{eq:upperbound}
\end{equation}

In order to obtain a first crude estimate, we may take advantage of some
trivial but nevertheless fundamental inequality. This inequality relates the
expectation values of both the first and second powers of any self-adjoint
but otherwise arbitrary operator ${\cal O} = {\cal O}^\dagger$ taken with
respect to arbitrary Hilbert-space vectors $|\rangle$ (in the domain of
${\cal O}$) normalized to unity; it reads
$$
|\langle{\cal O}\rangle| \le \sqrt{\langle{\cal O}^2\rangle} \ .
$$
Applying this inequality to the kinetic-energy part of the above expression
for $E(\lambda)$, we may replace, in turn, $E(\lambda)$ by an upper bound
which can be evaluated much easier than $E(\lambda)$ itself:
$$
E(\lambda) \le 2\sqrt{\langle\lambda|{\bf p}^2|\lambda\rangle + m^2}
+ \langle\lambda|V({\bf x})|\lambda\rangle \ .
$$
Identifying in this---as far as its evaluation is concerned,
simplified---upper bound the until-now general potential $V({\bf x})$ with
the Coulomb potential (\ref{eq:coulpot}) and inserting both of the
$\lambda$-dependent expectation values given above implies $$E(\lambda) \le
2\sqrt{\lambda^2 + m^2} - \kappa\,\lambda\ .$$

{}From this intermediate result, upon inspection of the limit
$\lambda\to\infty$, we may state already at this rather early stage that, for
the semirelativistic Hamiltonian (\ref{eq:semrelham}), (\ref{eq:coulpot}) to
be bounded from below at all, the Coulombic coupling strength $\kappa$ has to
stay below a certain critical value: $$\kappa \le 2\ .$$

The value of the variational parameter $\lambda$ which minimizes the latter
upper bound may be determined from the first derivative of this expression
with respect to $\lambda$:
$$
\lambda_{\rm min}
= \frac{m\,\kappa}{2\sqrt{1 - \displaystyle\frac{\kappa^2}{4}}} \ .
$$
For this particular value of $\lambda$, by shuffling together all our
previous inequalities, we find that the energy eigenvalue corresponding to
the ground state of the semirelativistic Hamiltonian (\ref{eq:semrelham})
with the Coulomb potential (\ref{eq:coulpot}), $E_0$, is bounded from above
by\cite{lucha94com}
$$
E_0 \le 2\,m\sqrt{1 - \frac{\kappa^2}{4}} \ .
$$
The reality of this upper bound requires again $\kappa \le 2$.

It is straightforward to improve this crude estimate by direct evaluation of
Eq.~(\ref{eq:upperbound}). For our choice of trial functions, the resulting
upper bound on the ground-state energy $E_0$ may be expressed in terms of the
hypergeometric function $F$:\cite{abramow}
$$
E_0 \le \left[\frac{128}{15\,\pi}\,
F\left(-\frac{1}{2},2;\frac{7}{2};1 - \frac{m^2}{\lambda^2}\right)
- \kappa\right] \lambda \ .
$$
Quite obviously, in this case the minimizing value of the variational
parameter $\lambda$ must be determined numerically. As before, the limit
$\lambda\to\infty$ tell us that now $$\kappa \le \frac{16}{3\,\pi}\ .$$

\section{Numerical Approximation Methods}

For an arbitrary interaction potential, it is, in general, not possible to
find a closed solution to the spinless Salpeter equation. Therefore, several
numerical approximation methods for the (numerical) solution of this equation
have been introduced.\cite{nickisch84,jacobs86,durand90,lucha91num,fulcher93}
All of these numerical schemes aim at the conversion of the spinless Salpeter
equation into an equivalent matrix eigenvalue problem, and confine themselves
to the case of a spherically symmetric potential $V(r)$. Here, we briefly
sketch the maybe most efficient among them, namely, the ``semianalytical
matrix method'' developed by Lucha {\it et~al.}\cite{lucha91num}

For states of definite orbital angular momentum $\ell$ described by some wave
function $\psi({\bf x})$, we define the reduced radial wave function $u(r)$
by
$$
\psi({\bf x}) = \frac{u(r)}{r}\;{\cal Y}_{\ell m}(\theta,\phi) \ ;
$$
here ${\cal Y}_{\ell m}(\theta,\phi)$ are the spherical harmonics of angular
momentum $\ell$ and projection $m$. For this reduced radial wave function
$u(r)$, an integral representation\cite{nickisch84} of the spinless Salpeter
equation may be found. Furthermore, since we consider only the case of equal
(nonvanishing) masses $m$ of the bound-state constituents, we may eliminate
from the kinetic-energy part of this integral equation any dependence on the
mass $m$ by scaling the radial variable $r$ like $$x := m\,r\ .$$ Introducing
the scaled reduced radial wave function $$\tilde u(x) :=
u\left(\frac{x}{m}\right) = u(r)$$ as well as the dimensionless energy
eigenvalue $$\tilde E := \frac{E}{m}$$ and the dimensionless interaction
potential $$\tilde V(x) := \frac{V\left(\frac{x}{m}\right)}{m} =
\frac{V(r)}{m}\ ,$$ this particular integral representation of the spinless
Salpeter equation becomes\cite{nickisch84}
\begin{equation}
\left[\tilde E - \tilde V(x)\right]\tilde u(x)
= \frac{2}{\pi}\int\limits_0^\infty dy\,G_\ell(x,y)
\left[- \frac{d^2}{dy^2} + \frac{\ell\,(\ell + 1)}{y^2} + 1\right]\tilde u(y)
\ ,
\label{eq:sseq-scales}
\end{equation}
where the kernel $G_\ell$ is defined by
$$
G_\ell(x,y) = 2^\ell\,z^{\ell + 1}
\left(\frac{1}{z}\,\frac{\partial}{\partial z}\right)^\ell \frac{1}{z}
\left[(s - z)^{\ell/2}\,K_\ell\left(\sqrt{s - z}\right)
- (s + z)^{\ell/2}\,K_\ell\left(\sqrt{s + z}\right)\right] \ ,
$$
with
\begin{eqnarray*}
s &\equiv& x^2 + y^2\ ,\\ z &\equiv& 2\,x\,y\ .
\end{eqnarray*}
Here $K_\ell$ is the modified Bessel function\cite{abramow} of the second
kind of order $\ell$.

In the free case, i.~e., for $$V(r) \equiv 0\ ,$$ as well as for potentials
which are less singular than the Coulomb potential, i.~e., $$V(r) \propto
\frac{1}{r^\eta}\quad\mbox{with}\ \eta < 1\ ,$$ the reduced radial wave
function $u(r)$ behaves for small $r$, $r \rightarrow 0$, asymptotically like
$$u(r) \propto r^{\ell + 1}\ .$$ Accordingly, we make the ansatz $$\tilde
u(x) = x^{\ell + 1}\,w(x)\ .$$ With this substitution, the
integro-differential equation (\ref{eq:sseq-scales}) is equivalent to
$$
\left[\tilde E - \tilde V(x)\right] x^{\ell + 1}\,w(x)
= \frac{2}{\pi}\int\limits_0^\infty dy\,G_\ell(x,y)\,y^{\ell + 1}
\left[1 - \frac{d^2}{dy^2} - \frac{2\,(\ell + 1)}{y}\,\frac{d}{dy}\right]
w(y) \ .
$$

In order to solve this scaled form of the spinless Salpeter equation, we
expand the solution $w(x)$ of this integro-differential equation into a
complete orthonormal system $\{f_n(x),\ n = 0,1,2,\dots\}$ of basis functions
for $L_2(R^3)$:
$$
w(x) = \sum_{n=0}^N \lambda_n\,f_n(x) \ ,
$$
with some set of real coefficients $\lambda_n$. For finite $N$, this
expansion represents, of course, only an approximation to the exact solution
$w(x)$. The crucial point\cite{lucha91num} of the present approach is our
rather sophisticated choice of the basis functions $f_n$ which, at least in
principle, allows for a thorough analytical treatment of the spinless
Salpeter equation: $$f_n(x) := \sqrt{2}\,\exp(-x)\,L_n(2x)\ ,$$ where
$L_n(x)$ are the Laguerre polynomials.\cite{abramow}

After some straightforward algebra, the scaled spinless Salpeter equation may
be cast into the form of a matrix eigenvalue equation for the coefficient
vector $\lambda \equiv \{\lambda_n\}$ in the above expansion of the solution
$w(x)$. In self-explanatory matrix notation, this eigenvalue equation is
given by\cite{lucha91num}
\begin{equation}
\tilde E\,\lambda = \left(P^{(\ell)}\right)^{-1}
\left[\left(T^{(\ell)}\right)^T + V^{(\ell)}\right] \lambda \ ,
\label{eq:numeigenequat}
\end{equation}
where the ``power matrix'' $P^{(\ell)}_{nm}$ and the ``potential matrix''
$V^{(\ell)}_{nm}$ are defined by
\begin{eqnarray*}
P^{(\ell)}_{nm} &:=& \int\limits_0^\infty dx\,x^{\ell + 1}\,f_n(x)\,f_m(x)
= P^{(\ell)}_{mn} \ ,\\
V^{(\ell)}_{nm} &:=&
\int\limits_0^\infty dx\,x^{\ell + 1}\,\tilde V(x)\,f_n(x)\,f_m(x)
= V^{(\ell)}_{mn} \ ,
\end{eqnarray*}
and the ``kinetic matrix'' $T^{(\ell)}_{nm}$ represents the action of the
kinetic term in the spinless Salpeter equation on the vector of basis
functions $f_n$. In this way, the solution of the spinless Salpeter equation
can be reduced to a simple matrix eigenvalue problem. The eigenvalues of this
equation are the energies $E$ of the bound state under consideration. The
corresponding eigenvectors $\{\lambda_n\}$ give the radial wave functions
$u(r)$ according to
$$
u(r) = (mr)^{\ell + 1} \sum_{n=0}^N \lambda_n\,f_n(mr) \ .
$$
For $N = \infty$ this treatment would be exact. For $N < \infty$ it provides
an approximation to the exact solution of increasing accuracy with increasing
$N$, that is, with increasing size of the involved matrices. As is evident
from the above construction, this procedure works analytically, at least for
all potentials of the type ``power times exponential,'' i.~e., for all those
potentials which involve only terms of the form $$r^n\exp(- b_n\,r)\ ,$$ with
(maybe vanishing) constants $b_n$.

Let us illustrate this simple prescription for the case of bound states with
vanishing orbital angular momentum $\ell$ of their constituents, that is,
$\ell = 0$. All we need are the explicit expressions of the kinetic matrix
$T^{(0)}$ and of the inverse of the power matrix $P^{(0)}$ for $\ell = 0$.
The first few entries $T^{(0)}_{nm}$ in the kinetic matrix $T^{(0)}$
read\cite{lucha91num}
$$
T^{(0)}_{nm} = \frac{2}{\pi}\,\frac{8}{(2\,n + 2\,m + 3)!!}\,S_{nm} \ ,
$$
where the matrix $S$ is given by
$$
S = \left(\begin{array}{rrrrr}
 1& -3& -5& -21&\cdots\\
 -1& 81& -375& -1029&\cdots\\
 -1& -225& 13125& -77175&\cdots\\
 -3& -441&-55125&3565485&\cdots\\
\vdots&\vdots&\vdots& \vdots&\ddots\end{array}\right) \ .
$$
(Note the range of the indices $n$ and $m$: $n,m = 0,1,\dots ,N$.) The
elements $P^{(0)}_{nm}$ of the power matrix $P^{(0)}$ read
explicitly\cite{lucha91num}
\begin{eqnarray*}
P^{(0)}_{nm} &\equiv&
\frac{1}{2} \int\limits_0^\infty dx\,x\,\exp(-x)\,L_n(x)\,L_m(x)\\
&=& \frac{1}{2}\left\{\begin{array}{ll}
2\,n + 1 &\ \mbox{for}\ m = n\\
- m &\ \mbox{for}\ m = n + 1\\
- n &\ \mbox{for}\ m = n - 1\\
0 &\ \mbox{else} \end{array}\right.\\[1ex]
&=& \frac{1}{2}\left(\begin{array}{rrrrr}
 1& -1& 0& 0&\cdots\\
 -1& 3& -2& 0&\cdots\\
 0& -2& 5& -3&\cdots\\
 0& 0& -3& 7&\cdots\\
\vdots&\vdots&\vdots&\vdots&\ddots\end{array}\right) \ .
\end{eqnarray*}
The inverse of this matrix, required for the matrix eigenvalue equation
(\ref{eq:numeigenequat}), depends explicitly on the size $N$ of the involved
matrices and is given by\cite{lucha91num}
$$
\begin{array}{ll}
\left(P^{(0)}\right)^{-1}_{nn} = \displaystyle\sum_{k=n}^N \frac{2}{k+1}
&\ \mbox{for}\ n = 0,1,\dots ,N \ ,\\
\left(P^{(0)}\right)^{-1}_{nm} = \left(P^{(0)}\right)^{-1}_{mn}
= \left(P^{(0)}\right)^{-1}_{nn}
&\ \mbox{for}\ n = 0,1,\dots ,N,\ m = 0,1,\dots ,n - 1 \ .
\end{array}
$$
For instance, for $N = 3$ this inverse reads
$$
\left(P^{(0)}\right)^{-1}_{4 \times 4} = \frac{1}{6}\left(\begin{array}{rrrrr}
25&13&7&3\\
13&13&7&3\\
 7& 7&7&3\\
 3& 3&3&3\end{array}\right) \ .
$$

Consider, for example, the so-called funnel potential $V_{\rm F}(r)$, which
depends on just two parameters, namely, on the Coulomb coupling constant
$\kappa$ and on the slope $a$ of the linear term: $V(r) = V_{\rm F}(r) \equiv
- \kappa/r + a\,r$. This funnel-shaped potential represents the
prototype\cite{lucha91} of all ``realistic,'' that is, phenomenologically
acceptable, ``QCD-inspired'' static interquark potentials proposed for the
description of hadrons as bound states of (constituent) quarks in the
framework of potential models, the quarks inside a hadron being bound by the
strong interactions arising from quantum chromodynamics. Now, according to
the above definition, the scaled form of the funnel potential, $\tilde V_{\rm
F}(x)$, reads $$\tilde V_{\rm F}(x) = - \frac{\kappa}{x} + \frac{a}{m^2}\,x\
,$$ which entails for the corresponding potential matrix
$$
V_{\rm F}^{(\ell)}
= - \kappa\,P^{(\ell - 1)} + \frac{a}{m^2}\,P^{(\ell + 1)} \ .
$$
In particular, specializing again to the case $\ell = 0$, the matrix elements
of the funnel potential, taken with respect to $\ell = 0$ states, are given
by
$$
V_{\rm F}^{(0)} = - \kappa\,P^{(- 1)} + \frac{a}{m^2}\,P^{(1)} \ .
$$
Since, according to the orthonormalization condition for the basis functions
$\{f_n(x)\}$, the power matrix $P^{(- 1)}$ is identical to the unit matrix,
$$P^{(- 1)} = 1_{N \times N}\ ,$$ we have
$$
V_{\rm F}^{(0)} = - \kappa + \frac{a}{m^2}\,P^{(1)} \ .
$$
Working out in detail the $\ell = 1$ power matrix $P_{nm}^{(1)}$, we
obtain\cite{lucha91num}
$$
\begin{array}{ll}
P_{nn}^{(1)} = \displaystyle\frac{3\,n\,(n + 1) + 1}{2}
&\ \mbox{for}\ n = 0,1,\dots ,N \ ,\\
P_{n,n+1}^{(1)} = - n\,(n + 2) - 1
&\ \mbox{for}\ n = 0,1,\dots ,N - 1 \ ,\\
P_{n,n+2}^{(1)} = \displaystyle\frac{n\,(n + 3) + 2}{4}
&\ \mbox{for}\ n = 0,1,\dots ,N - 2 \ ,\\
P_{nm}^{(1)} = 0 &\ \mbox{else} \ .
\end{array}
$$
Explicitly, the first entries of $P^{(1)}$ read
$$
P^{(1)} = \frac{1}{2}\left(\begin{array}{rrrrr}
 1& -2& 1& 0&\cdots\\
 -2& 7& -8& 3&\cdots\\
 1& -8& 19& -18&\cdots\\
 0& 3& -18& 37&\cdots\\
\vdots&\vdots&\vdots&\vdots&\ddots\end{array}\right) \ .
$$

In summary, shuffling together the above results
\begin{itemize}
\item for the kinetic matrix $T^{(0)}$,
\item for the inverse of the power matrix $P^{(0)}$, and
\item for the potential matrix $V^{(0)}$ of the potential under consideration
\end{itemize}
one ends up with a well-defined matrix eigenvalue equation, which represents
the $\ell = 0$ special case of the general matrix form
(\ref{eq:numeigenequat}) of the spinless Salpeter equation.

\section{Effectively Semirelativistic Hamiltonians of Nonrelativistic Form}

Almost all of the troubles which one encounters when trying to apply the
spinless Salpeter equation are obviously brought about by the nonlocality of
the ``square-root'' operator of the relativistic kinetic energy, $\sqrt{{\bf
p}^2 + m^2}$, in the Hamiltonian $H$, Eq.~(\ref{eq:semrelham}). In contrast
to the nonrelativistic limit, obtained from the expansion of the square root
up to the lowest ${\bf p}^2$-dependent order, $\sqrt{{\bf p}^2 + m^2} = m +
{\bf p}^2/(2\,m) + \dots$, the presence of this relativistic kinetic-energy
operator prevents, in general, a thoroughly analytic discussion; one is
forced to rely on numerical solutions of the problem. This (intrinsic)
difficulty of any (semi-)relativistic formalism may be circumvented by
approximating the Hamiltonian (\ref{eq:semrelham}) by the corresponding
``effectively semirelativistic'' Hamiltonian,\cite{lucha93eff} which retains
apparently the easier-to-handle nonrelativistic kinematics but resembles its
relativistic counterpart to the utmost possible extent by replacing some of
its basic parameters by effective ones which depend, in a well-defined
manner, on the square of the relevant momentum ${\bf p}$. The main
idea\cite{lucha91,lucha92} of the construction\cite{lucha93eff} of these
``effectively semirelativistic'' Hamiltonians is as follows.

Applying again the above-mentioned fundamental inequality
$$
|\langle{\cal O}\rangle| \le \sqrt{\langle{\cal O}^2\rangle} \ ,
$$
which holds for any self-adjoint operator ${\cal O} = {\cal O}^\dagger$ and
arbitrary Hilbert-space vectors $|\rangle$ (in the domain of ${\cal O}$)
normalized to unity, to the relativistic kinetic-energy operator $\sqrt{{\bf
p}^2 + m^2}$ yields
$$
\left\langle\sqrt{{\bf p}^2 + m^2}\right\rangle
\le \sqrt{\langle{\bf p}^2\rangle + m^2} \ .
$$
By employing this inequality, we obtain for an arbitrary expectation value
$\langle H\rangle$ of the semirelativistic Hamiltonian $H$,
Eq.~(\ref{eq:semrelham}),
\begin{eqnarray}
\langle H\rangle
&=& 2\left\langle\sqrt{{\bf p}^2 + m^2}\right\rangle
+ \langle V\rangle\nonumber\\
&\le& 2\,\sqrt{\langle{\bf p}^2\rangle + m^2} + \langle V\rangle\nonumber\\
&=& 2\,\frac{\langle{\bf p}^2\rangle + m^2}
{\sqrt{\langle{\bf p}^2\rangle + m^2}} + \langle V\rangle\nonumber\\
&=& \left\langle 2\,\frac{{\bf p}^2 + m^2}
{\sqrt{\langle{\bf p}^2\rangle + m^2}} + V\right\rangle \ .
\label{eq:hamexpectval}
\end{eqnarray}

{}From now on we specify the generic Hilbert-space vectors in all expectation
values to be the eigenstates of our Hamiltonian $H$. In this case the
expectation value of $H$, $\langle H\rangle$, as appearing, e.~g., in
Eq.~(\ref{eq:hamexpectval}), becomes the corresponding semirelativistic
energy eigenvalue $E$, i.~e., $$E \equiv \langle H\rangle\ ,$$ and the
inequality (\ref{eq:hamexpectval}) tells us that this energy eigenvalue is
bounded from above by\cite{lucha91,lucha92}
$$
E \le \left\langle 2\,\frac{{\bf p}^2 + m^2}
{\sqrt{\langle{\bf p}^2\rangle + m^2}} + V\right\rangle \ .
$$
The operator within brackets on the right-hand side of this inequality may be
regarded as some ``effectively semirelativistic'' Hamiltonian $H_{\rm eff}$
which possesses, quite formally, the structure of a nonrelativistic
Hamiltonian,\cite{lucha91,lucha92}
\begin{eqnarray}
H_{\rm eff}
&\equiv& 2\,\frac{{\bf p}^2 + m^2}{\sqrt{\langle{\bf p}^2\rangle + m^2}} + V
\nonumber\\
&=& 2\,\hat m + \frac{{\bf p}^2}{\hat m} + V_{\rm eff} \ ,
\label{eq:heffective}
\end{eqnarray}
but involves, however, the effective mass\cite{lucha91,lucha92}
\begin{equation}
\hat m = \frac{1}{2}{\sqrt{\langle{\bf p}^2\rangle + m^2}}
\label{eq:meffective}
\end{equation}
as well as the effective nonrelativistic potential\cite{lucha91,lucha92}
\begin{eqnarray}
V_{\rm eff} &=& \frac{2\,m^2}{\sqrt{\langle{\bf p}^2\rangle + m^2}}
- \sqrt{\langle{\bf p}^2\rangle + m^2} + V\nonumber\\
&=& 2\,\hat m - \frac{\langle{\bf p}^2\rangle}{\hat m} + V \ .
\label{eq:poteffective}
\end{eqnarray}
The effective mass $\hat m$ given by Eq.~(\ref{eq:meffective}) and the
constant, i.~e., coordinate-independent, term in the effective potential
$V_{\rm eff}$ of Eq.~(\ref{eq:poteffective}), $$2\,\hat m - \frac{\langle{\bf
p}^2\rangle}{\hat m}\ ,$$ obviously depend on the expectation value of the
square of the momentum ${\bf p}$, that is, on $\langle{\bf p}^2\rangle$, and
will therefore differ for different energy eigenstates.

Motivated by our above considerations, we propose to approximate the true
energy eigenvalues $E$ of the semirelativistic Hamiltonian $H$ of
Eq.~(\ref{eq:semrelham}) by the corresponding ``effective'' energy
eigenvalues $E_{\rm eff}$, defined as the expectation values of some
effective Hamiltonian $\tilde H_{\rm eff}$ taken with respect to the
eigenstates $|\rangle_{\rm eff}$ of its own, $$E_{\rm eff} = \langle\tilde
H_{\rm eff}\rangle_{\rm eff}\ ,$$ where the effective Hamiltonian $\tilde
H_{\rm eff}$, as far as its structure is concerned, is given by
Eqs.~(\ref{eq:heffective}) through (\ref{eq:poteffective}) but is implicitly
understood to involve the expectation values of ${\bf p}^2$ with respect to
the effective eigenstates $|\rangle_{\rm eff}$ (that is, $\langle{\bf
p}^2\rangle_{\rm eff}$ in place of $\langle{\bf
p}^2\rangle$):\cite{lucha93eff}
$$
\tilde H_{\rm eff} = 4\,\tilde m
+ \frac{{\bf p}^2 - \langle{\bf p}^2\rangle_{\rm eff}}{\tilde m} + V\ ,
$$
with
$$
\tilde m = \frac{1}{2}{\sqrt{\langle{\bf p}^2\rangle_{\rm eff} + m^2}} \ .
$$
Accordingly, the effective energy eigenvalues $E_{\rm eff}$ are given by a
rather simple formal expression, viz., by\cite{lucha93eff}
\begin{equation}
E_{\rm eff} = 4\,\tilde m + \langle V\rangle_{\rm eff} \ .
\label{eq:effenergy}
\end{equation}

We intend to elaborate our general prescription for the construction of
effectively semirelativistic Hamiltonians $\tilde H_{\rm eff}$ in more detail
for the particular case of power-law potentials depending only on the radial
coordinate $r \equiv |{\bf x}|$, i.~e., for potentials of the form $$V(r) =
a\,r^n$$ with some constant $a$. The reason for this restriction is twofold:
\begin{enumerate}
\item On the one hand, for power-law potentials the general virial
theorem\cite{lucha89,lucha90mpla} in its nonrelativistic
form\cite{lucha91,lucha92} appropriate for the present case,
$$
\left\langle\frac{{\bf p}^2}{\tilde m}\right\rangle_{\rm eff}
= \frac{1}{2}\left\langle r\,\frac{dV(r)}{dr}\right\rangle_{\rm eff} \ ,
\label{eq:virial}
$$
enables us to replace the expectation value of the potential in
(\ref{eq:effenergy}) immediately by a well-defined function of the
expectation value of the squared momentum:
$$
a\,\langle r^n\rangle_{\rm eff}
= \frac{2}{n}\,\frac{\langle{\bf p}^2\rangle_{\rm eff}}{\tilde m} \ .
$$
This implies for the effective energy eigenvalues
\begin{equation}
E_{\rm eff} = 4\,\tilde m + \frac{2}{n}\,
\frac{\langle{\bf p}^2\rangle_{\rm eff}}{\tilde m} \ .
\label{eq:eeff}
\end{equation}
\item On the other hand, we take advantage of the fact that for power-law
potentials it is possible to pass, without change of the fundamental
commutation relations between coordinate variables and their canonically
conjugated momenta, from the dimensional phase-space variables adopted at
present to new, dimensionless phase-space variables and to rewrite the
Hamiltonian in form of a Hamiltonian which involves only these dimensionless
phase-space variables.\cite{lucha91} The eigenvalues $\epsilon$ of this
dimensionless Hamiltonian are, of course, also dimensionless.\cite{lucha91}
Applying this procedure, we find for the effective energy eigenvalues
\begin{eqnarray*}
E_{\rm eff} - 4\,\tilde m
+ \frac{\langle {\bf p}^2\rangle_{\rm eff}}{\tilde m}
&=& \left\langle\frac{{\bf p}^2}{\tilde m} + a\,r^n\right\rangle_{\rm eff}\\
&=& \left(\frac{a^2}{{\tilde m}^n}\right)^{\frac{1}{2+n}} \epsilon \ .
\end{eqnarray*}
\end{enumerate}
Combining both of the above expressions for $E_{\rm eff}$, we obtain a
relation which allows us to determine $\langle{\bf p}^2\rangle_{\rm eff}$
unambiguously in terms of the dimensionless energy eigenvalues
$\epsilon$:\cite{lucha93eff}
\begin{equation}
\langle{\bf p}^2\rangle_{\rm eff}^{2+n}
= \frac{1}{4}\left(\frac{n}{2+n}\right)^{2+n}a^2\,\epsilon^{2+n}\,
(\langle{\bf p}^2\rangle_{\rm eff} + m^2) \ .
\label{eq:master}
\end{equation}
For a given power $n$ this equation may be solved for $\langle{\bf
p}^2\rangle_{\rm eff}$. Insertion of the resulting expression into
Eq.~(\ref{eq:eeff}) then yields the corresponding eigenvalue $E_{\rm eff}$ of
the effectively semirelativistic Hamiltonian $\tilde H_{\rm eff}$.

Bound-state solutions are usually characterized by some radial quantum number
$n_{\rm r}$ and orbital angular-momentum quantum number $\ell$. For instance,
for the harmonic oscillator, i.~e., for $n = 2$, the dimensionless energy
eigenvalues $\epsilon$ are given by\cite{lucha91} $$\epsilon = 2\,N$$ with
$$N = 2\,n_{\rm r} + \ell + \frac{3}{2}\ .$$

For the Coulomb potential $V(r) = - \kappa/r$, that is, for $n = - 1$,
Eq.~(\ref{eq:master}) reduces to a linear equation for the expectation value
$\langle{\bf p}^2\rangle_{\rm eff}$. Inserting the well-known
expression\cite{lucha91} for the dimensionless energy eigenvalues $\epsilon$
of the (nonrelativistic) Coulomb problem, $$\epsilon = - \frac{1}{(2\,N)^2}$$
with $$N = n_{\rm r} + \ell + 1\ ,$$ we obtain from this linear equation for
$\langle{\bf p}^2\rangle_{\rm eff}$
$$
\langle{\bf p}^2\rangle_{\rm eff}
= \frac{\kappa^2\,m^2}{16\,N^2 - \kappa^2} \ ,
$$
and, after inserting this expression into Eq.~(\ref{eq:eeff}), for the
effective energy eigenvalue\cite{lucha93eff}
$$
E_{\rm eff}
= \frac{m}{N}\,\frac{8\,N^2 - \kappa^2}{\sqrt{16\,N^2 - \kappa^2}} \ .
$$

The capability of this effective formalism to imitate the semirelativistic
treatment becomes apparent by inspecting, for instance, the behaviour of the
energy eigenvalues $E_{\rm eff}$ for large orbital angular momenta $\ell$.
For the linear potential $V(r) = a\,r$, that is, for $n = 1$, the effective
formalism yields (in the ultrarelativistic limit, i.~e., for $m=0$) linear
Regge trajectories:\cite{lucha93eff} $$E_{\rm eff}^2 = 9\,a\,\ell\ .$$ This
fits very well to the exact semirelativistic
result,\cite{kang75,lucha91regge} $$E_{\rm SR}^2 = 8\,a\,\ell\ ,$$ but is in
clear contrast to the nonrelativistic approach, which
gives\cite{lucha91,lucha92} $$E_{\rm NR}^2 =
9\left(\frac{a^2}{4\,m}\right)^{2/3}\ell^{4/3}\ .$$

\section{Summary and Conclusion}

In spite of the regrettable fact that, at present, not even for the Coulomb
potential the exact solution for the ground-state energy is known, the
spinless Salpeter equation is certainly a useful tool for the
semirelativistic description of bound states consisting of scalar bosons only
as well as of the spin-averaged spectra of bound states consisting of
fermionic constituents.

\vspace{0.6cm}
\leftline{\bf Acknowledgements}
\vspace{0.4cm}
We would like to thank J.~Ball, L.~P.~Fulcher, and I.~W.~Herbst for some
interesting discussions.

\end{document}